# Regulation of star formation in giant galaxies by precipitation, feedback, and conduction

G. M. Voit[1], M. Donahue[1], G. L. Bryan[2], M. McDonald[3]

[1]Department of Physics and Astronomy, 567 Wilson Road, Michigan State University, East Lansing, MI 48864, USA
[2]Department of Astronomy, 1328 Pupin Physics Lab, MC 5246, 550 West 120th Street, New York, NY 10027, USA
[3]MIT Kavli Institute for Astrophysics and Space Research, MIT, 77 Massachusetts Avenue, Cambridge, MA 02139, USA

**The universe's largest galaxies reside at the centers of galaxy clusters and are embedded in hot gas that, if left unchecked, would cool prodigiously and create many more new stars than are actually observed.[1–5] Cooling can be regulated by feedback from accretion of cooling gas onto the central black hole, but requires an accretion rate finely tuned to the thermodynamic state of the hot gas.[6,7] Theoretical models in which cold clouds precipitate out of the hot gas via thermal instability and accrete onto the black hole exhibit the necessary tuning.[8–10] We have recently presented observational evidence showing that the abundance of cold gas in the central galaxy increases rapidly near the predicted threshold for instability.[11] Here we present observations showing that this threshold extends over a large range in cluster radius, cluster mass, and cosmic time, and incorporate the precipitation threshold into a comprehensive framework of theoretical models for the thermodynamic state of hot gas in galaxy clusters. According to that framework, precipitation regulates star formation in some giant galaxies, while thermal conduction prevents star formation in others, if it can compensate for radiative cooling and shut off precipitation.**

Our framework can be expressed in terms of the time $t_{\rm cool}$ required for the hot gas to radiate an amount of energy equivalent to its current thermal energy. If intracluster gas were unable to cool, cosmological structure formation via hierarchical merging would produce galaxy clusters with radial cooling-time profiles similar to a baseline profile $t_{\rm base}(r)$ that can be computed with numerical simulations.[12,13] Massive galaxy clusters are observed to converge to this baseline profile at large radii,[14] but radiative cooling cannot be ignored at smaller radii, where $t_{\rm cool}$ can be much shorter than the age of the universe. Gas at small radii must either cool and condense or cooling of that gas must trigger thermal feedback that compensates for the radiative losses.[15]

Thermal conduction is capable of compensating for cooling in cluster gas with $t_{\rm cool} > 1$ Gyr.[16,17] Our framework therefore includes a locus of conductive balance, $t_{\rm cond}(r)$, along which thermal conduction exactly balances radiative cooling.[18] The locus itself is unstable, because conduction outcompetes cooling if $t_{\rm cool}$ is above that locus but cannot compete below it.[19] Conduction should therefore drive gas above the locus toward an isothermal core profile $t_{\rm iso}(r)$ identical to the baseline profile at large radii but with a constant temperature equal to the peak temperature of the baseline profile at smaller radii. Clusters in an isothermal core state have central cooling times exceeding ~1 Gyr, and so mergers with other galaxy clusters, which occur on timescales of several Gyr, can compete with cooling and further raise $t_{\rm cool}$ in the cores of those objects. Once $t_{\rm cool}$ exceeds the 14 Gyr age of the universe, radiative cooling can no longer lower $t_{\rm cool}$, and this threshold corresponds to the "no cooling" profile in our framework.

Clusters with cooling-time profiles that go below the locus of conductive balance require another heat source to balance cooling, and observations have shown that outflows emanating from a central supermassive black hole are sufficiently energetic to stop the cooling.[7] However, the triggering mechanism for that feedback response remained elusive until recent numerical simulations provided the missing puzzle piece.[8-10,20,21] Those simulations show that cold clouds start to precipitate out of hot-gas atmospheres in a state of global thermal balance when $t_{cool}$ drops to 10 times the free-fall time $t_{ff} = (2r/g)^{1/2}$, where $g$ is the local gravitational acceleration. The resulting precipitation feeds the central black hole through a "chaotic cold accretion" process, producing a combination of thermal and kinetic feedback that maintains the necessary state of global balance.[22,23] Sporadic eruptions of feedback then cause the minimum value of $t_{cool}/t_{ff}$ to fluctuate within the range $5 < t_{cool}/t_{ff} < 20$.

We compute the critical profile for precipitation[24] by assuming a two-component gravitational potential. The first component is a mass-density profile $\propto (3\,r/r_{500})^{-1}[1 + (3r/r_{500})]^{-2}$ in which the mean density within $r_{500}$ is 500 times the cosmological critical density,[25] and $r_{500}$ depends on the cluster's gas temperature via $kT_X \approx 125\mu m_p [H(z)\,r_{500}]^2$, where $H(z)$ is the Hubble expansion parameter at the cluster's cosmological redshift $z$ and $\mu m_p$ is the mean mass per gas particle. The second component is a singular isothermal sphere ($\rho \propto r^{-2}$) with a velocity dispersion of 250 km s$^{-1}$ to represent the stellar mass profile of the central galaxy.[26] Defining $t_{precip}(r) = 10 t_{ff}$ then yields the critical profile for precipitation of cold clouds out of the hot gas.

Notice that there are then three "attractor" profiles for cluster cores: (1) dynamical heating via mergers will push hot gas toward a long-lived state with $t_{cool} > 14$ Gyr, (2) thermal conduction will drive hot gas above the conductive-balance locus toward $t_{iso}(r)$, and (3) hot gas below the conductive-balance locus will cool, sink into the central galaxy, fall into a precipitating state, and trigger feedback that prevents $t_{cool}$ from dropping much below $10 t_{ff}$.

Comparing this framework of models with cooling-time profiles derived from the *ACCEPT* galaxy-cluster database[14] strongly supports the hypothesis that precipitation regulates cooling and star formation in massive galaxies (Figure 1). The lower envelope of the $t_{cool}(r)$ data closely follows the max[$t_{precip}(r), t_{base}(r)$] boundary over multiple orders of magnitude in radius, multiple orders of magnitude in cooling time, and more than an order of magnitude in system temperature. It even reproduces the kink at the intersection of $t_{precip}(r)$ and $t_{base}(r)$, confirming that the mechanism which regulates cooling and star formation in the universe's largest galaxies prevents $t_{cool}$ from dropping much below $10 t_{ff}$. This is an important finding, even if the precipitation-driven feedback model turns out to be incorrect, because it shows that the mechanism preventing runaway cooling in cluster cores depends critically on the $t_{cool}/t_{ff}$ ratio.

The data also imply that thermal conduction separates precipitating clusters from non-precipitating clusters because the locus of unstable conductive balance neatly divides systems with multiphase gas from those without it. Detections of Hα and far-infrared emission from cluster cores[14,27] indicate the presence of multiphase gas, and the cooling-time profiles of all multiphase cluster cores either drop below $t_{cond}(r)$ or are in its vicinity. In contrast, nearly all of the clusters without observable Hα emission stay above $t_{cond}(r)$. The few single-phase cluster cores that dip below $t_{cond}(r)$ may be objects in transition to a precipitating state because they are still outside the precipitation zone at $5 < t_{cool}/t_{ff} < 20$. According to our framework, their multiphase counterparts with $20t_{ff} < t_{cool} < t_{cond}(r)$ are likely to be systems in which a large burst of feedback has temporarily shut off precipitation but has not yet boosted $t_{cool}$ high enough for conduction to eliminate the multiphase gas. Those cluster cores should cool and return to active precipitation within a few hundred Myr.

Subdividing clusters according to temperature strengthens the case for thermal conduction (Figure 2). Our framework predicts that clusters with cooling-time profiles between the conductive-balance locus and the isothermal-core profile should be rare, because thermal conduction should be driving $t_{cool}$ from $t_{cond}(r)$ toward $t_{iso}(r)$. The data show that the zone between $t_{cond}(r)$ and $t_{iso}(r)$ is indeed systematically depopulated and suggest that it grows larger with increasing temperature, in accord with the strong temperature dependence of thermal conduction in astrophysical plasmas. Notice also that in every temperature range, the lower edge of the $t_{cool}$ envelope closely follows the joint precipitation+baseline profile, including the kink at the intersection point, showing that the floor at $t_{cool} \approx 10t_{ff}$ is present in data across the entire cluster temperature range.

Two predictions for the evolution of galaxy-cluster cores follow from these considerations. First, the thermodynamic properties of precipitating cores should remain relatively constant with time, because they are determined by local conditions and not by cosmological evolution. Second, the contrast in gas density between a precipitating core and the outer parts of a cluster should grow more pronounced with time, because hierarchical structure formation causes the baseline profile to become less dense as dynamical heating resulting from mergers adds entropy to the gas and shifts $t_{cool}$ upward. X-ray observations of the South Pole Telescope galaxy-cluster sample[28] support these predictions (Figure 3). The limits on central density, entropy, and cooling time of high-redshift clusters remain similar to those for low-redshift clusters and do not violate the precipitation limit, whereas the outer parts remain limited by the baseline profile, which is at progressively greater density, lower entropy, and shorter cooling time as cluster redshift increases.

Taken as a whole, this many-faceted correspondence between models and data convincingly shows that we now understand what regulates cooling and star formation in the universe's largest galaxies and raises an even bigger question: How far down the galaxy-mass spectrum do these principles extend? Precipitation is likely to be a very general feature of galaxy evolution, in that precipitation-driven feedback owing to both star formation and accretion onto black holes is likely to maintain the ambient circumgalactic medium of a star-forming galaxy in a state with $t_{cool}/t_{ff} \approx 10$. Conversely, galaxies embedded in ambient gas with $t_{cool}/t_{ff} \gg 10$ have no way of replenishing the cold gas required for star formation, which therefore wanes. Thermal conduction is probably less general, given its strong temperature dependence, but stellar heating mechanisms such as supernova explosions should be of greater relative importance in lower-temperature systems and may provide an analogous upper bound on residual precipitation that separates star-forming galaxies from those in which star formation has ceased.

**Acknowledgements** G.M.V. and M.D. acknowledge NSF support through grant AST-0908819. G.L.B. acknowledges NSF AST-1008134, AST-1210890, NASA grant NNX12AH41G, and XSEDE Computational resources. M. McD. acknowledges support by NASA through a Hubble Fellowship grant HST-HF51308.01-A awarded by the Space Telescope Science Institute, which is operated by the Association of Universities for Research in Astronomy, Inc., for NASA, under contract NAS 5-26555.

**Author Contributions**  G.M.V.: theoretical models, data interpretation, manuscript preparation; M.D.: data analysis, data interpretation, discussions, manuscript review; G.L.B.: theoretical models, discussions, manuscript review; M.McD.: data analysis, discussions, manuscript preparation, manuscript review

**Author Information** Reprints and permissions information is available at www.nature.com/reprints. The authors declare no competing financial interests. Readers are welcome to comment on the online version of the paper. Correspondence and requests for materials should be addressed to G.M.V. (voit@pa.msu.edu)

# Figure Captions

**Figure 1 | Hot-gas cooling time as a function of radius in galaxy clusters.** The observed ratio of cooling time to freefall time exhibits a hard floor at ≈ 10, in accordance with model predictions[24] for precipitation-driven feedback. Dashed blue lines show cooling-time profiles for all objects in the *ACCEPT* database[14] with gas temperatures in the 2–10 keV range and Hα detections of multiphase gas. Solid purple lines show all 0.5–2.0 keV objects in *ACCEPT* with far-infrared detections of multiphase gas. The lower envelope of the cooling-time profiles closely follows the boundary defined by the precipitation threshold at $t_{cool}/t_{ff}$ ≈ 10 (thick magenta line) and the cosmological baseline profile (brown), and most of those profiles enter the (pink) zone at 5 < $t_{cool}/t_{ff}$ < 20, within which precipitation-driven feedback stabilizes simulated galaxy clusters. The upper end of the $t_{cool}$ envelope for multiphase systems lies in the vicinity of the (cyan) locus of unstable conductive balance, indicating that thermal conduction eliminates multiphase gas above that locus. Dashed red lines show cooling-time profiles for all 2–10 keV objects in the *ACCEPT* database and no observable Hα emission. None of those profiles enters the precipitation zone, nearly all are above the locus of conductive balance, and most are between the (green) isothermal core profile and the (orange) cooling threshold at which the minimum cooling time equals the age of the universe. All of the thick solid lines show model predictions for a 6 keV cluster, and purple tags indicate the core entropy index ($K_0$ in keV cm$^2$) at this temperature. An error bar near the upper right corner shows the typical uncertainty range (2 times the s.d.) for $t_{cool}$, which comes primarily from the statistical uncertainty in gas temperatures derived from *Chandra* X-ray spectroscopy.

**Figure 2 | Hot-gas cooling time as a function of radius in galaxy clusters of differing temperatures.** All lines are colour-coded as in Figure 1. When grouped by temperature, all of the Hα–emitting clusters have profiles that dip below the locus of conductive balance, while only 3 of the no-Hα clusters dip below it. None of those three enters the pink range corresponding to the $t_{cool}/t_{ff}$ excursions seen in simulations of precipitation-driven feedback, suggesting they may be objects in which precipitation has not yet begun. In the yellow regions, our model predicts that thermal conduction should be heating gas and driving it to the isothermal-core state. If thermal conduction is indeed responsible for separating the $t_{cool}$ profiles of Hα and no-Hα clusters, then the degree of separation should increase with temperature. The main effect of increasing temperature is to drive the locus of conductive balance closer to the precipitation threshold, narrowing the range of $t_{cool}/t_{ff}$ within which multiphase gas can persist. This trend appears to be present in the data but with marginal statistical significance. For Hα–emitting clusters in the 2–7 keV range, we find that the mean value of min[$t_{cool}/t_{ff}$] is 20.9 ± 1.7 with a standard deviation of 9.5. Among Hα–emitting clusters in the 7–15 keV range, both the average value of min[$t_{cool}/t_{ff}$] and the dispersion are lower, with a mean of 15.7 ± 1.7 and a standard deviation of 5.6.

**Figure 3 | Evolution of radial gas-density and cooling-time profiles in galaxy clusters. a,** Evolution of electron density ($n_e$). Objects at mean cosmological redshift <z> = 0.12 are from the Chandra Cluster Cosmology Project,[29] and objects at <z> = 0.5 and 1.0 are from the South Pole Telescope[28] (SPT) survey. Cosmological scaling has been removed through division of $r$ by $r_{500}$ and division of $n_e$ by $\rho_{cr} f_b m_p^{-1}$, where $\rho_{cr}$ is the cosmological critical density and $f_b$ is the fraction

of cosmic mass in baryonic form. Thin lines show cluster observations. Solid thick magenta and brown lines show the precipitation limit and baseline profile, respectively, corresponding to a reference temperature of 6 keV. Dashed lines show the precipitation and baseline profiles for the low-redshift subsample at $<z> = 0.12$. The gas-density contrast between a core near the precipitation limit and the outer part of the baseline profile decreases with increasing redshift. This happens because the universe as a whole is denser at earlier times, whereas gas density at the precipitation limit remains nearly constant because it is set by local conditions. **b,** Evolution of hot-gas cooling time. All line styles are identical to those in panel (a). In this unscaled representation of the same data, the precipitation limit remains nearly constant, while the baseline profile shifts downward with increasing redshift because the mean gas density is increasing. Error bars in both panels show a statistical uncertainty range equivalent to 2 times the s.d. One SPT cluster in the $<z> = 0.5$ set, shown with a red line, crosses the precipitation limit. Notably, it is the Phoenix cluster,[30] which has, by far, the largest central star formation rate among all known galaxy clusters.

Figure 1

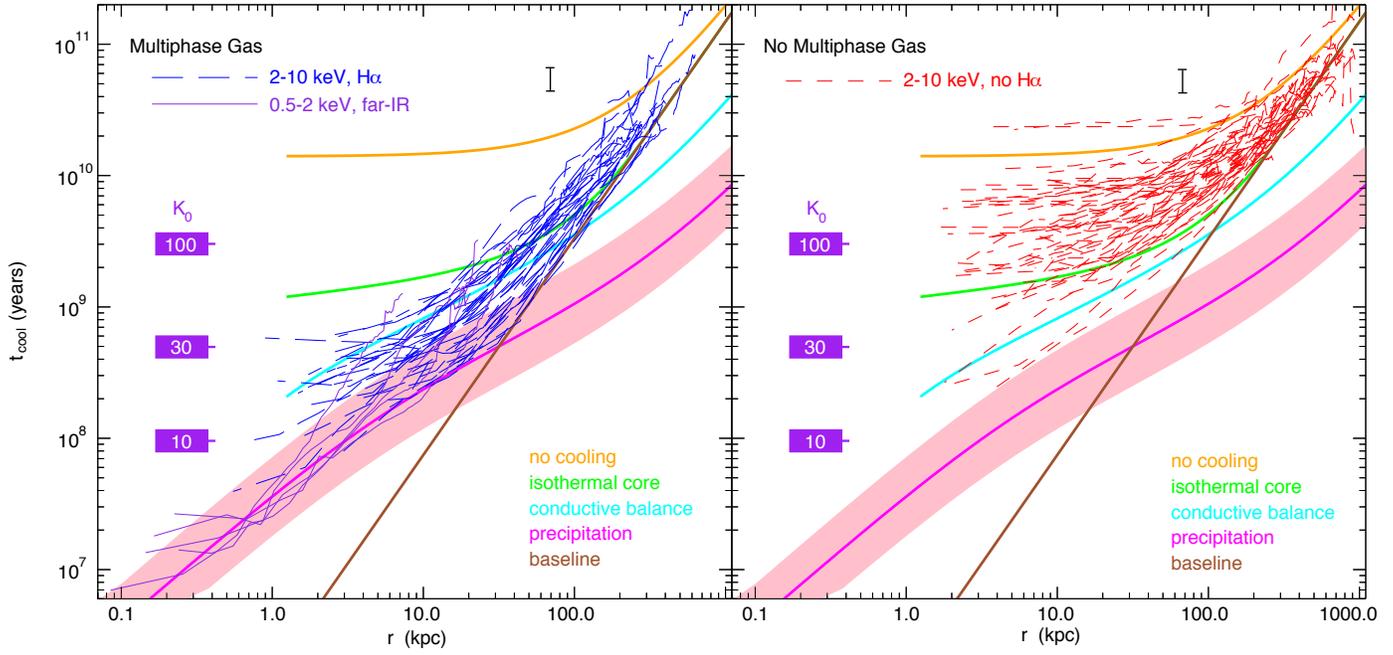

Figure 2

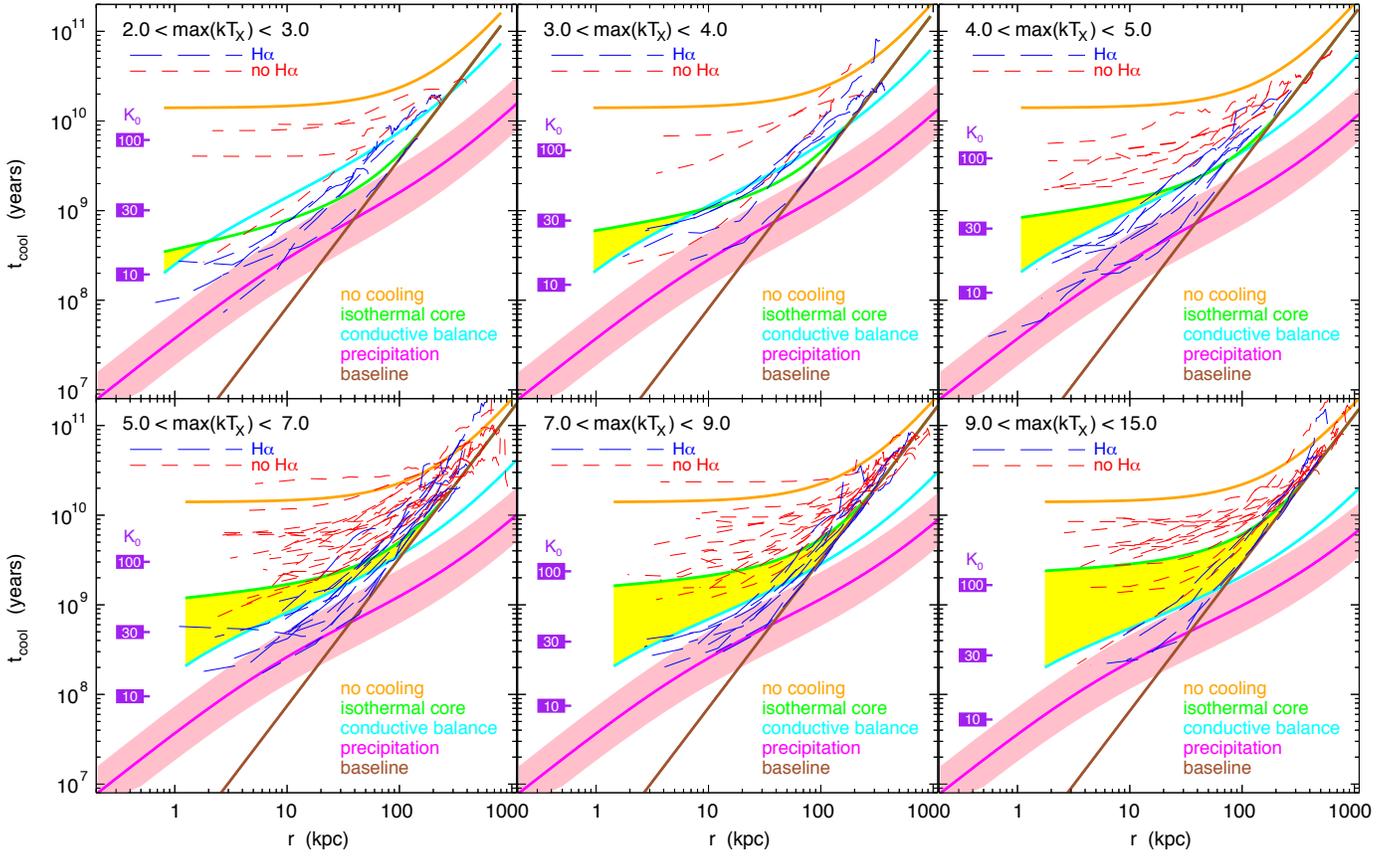

Figure 3

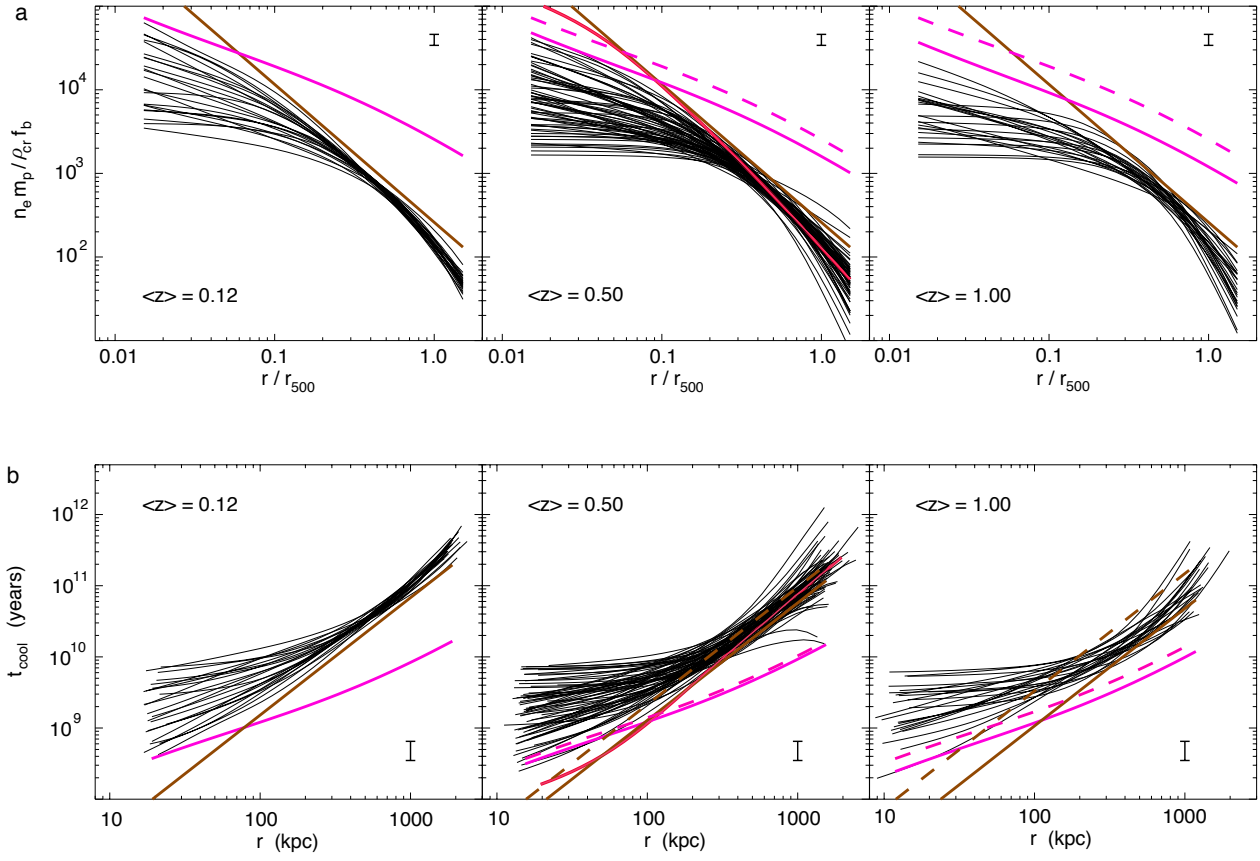